\documentclass[aps,onecolumn,letterpaper,oneside,preprint,tightenlines,draft,%
nobibnotes,nofootinbib,amsfonts,amssymb,amsmath,eqsecnum,showpacs,%
showkeys]{revtex4}
\usepackage{multirow}

\newcommand{\cD}{\mathcal{D}}
\newcommand{\cN}{\mathcal{N}}
\newcommand{\cP}{\mathcal{P}}
\newcommand{\cV}{\mathcal{V}}
\newcommand{\cW}{\mathcal{W}}
\newcommand{\bH}{\boldsymbol{H}}
\newcommand{\bQ}{\boldsymbol{Q}}
\newcommand{\tA}{\tilde{A}}
\newcommand{\tB}{\tilde{B}}
\newcommand{\tC}{\tilde{C}}
\newcommand{\tH}{\tilde{H}}
\newcommand{\tL}{\tilde{L}} 
\newcommand{\tP}{\tilde{P}}

\newcommand{\tcV}{\tilde{\cV}}
\newcommand{\tw}{\tilde{w}}
\newcommand{\tvph}{\tilde{\varphi}}

\newcommand{\bbN}{\mathbb{N}}

\newcommand{\bbR}{\mathbb{R}}
\newcommand{\bbC}{\mathbb{C}}
\newcommand{\rme}{\mathrm{e}}
\newcommand{\rmd}{\mathrm{d}}
\newcommand{\del}{\partial}

\newcommand{\rnu}{\sqrt{\nu}}
\newcommand{\braket}[1]{\bigl\langle{#1}\bigr\rangle}
\newcommand{\rmA}{\mathrm{A}}
\newcommand{\rmT}{\mathrm{T}}
\newcommand{\rmX}{\mathrm{X}}
\newcommand{\gd}{\operatorname{gd}}

\begin{document}


%
%

\title{$\cN$-fold Supersymmetry and Quasi-solvability Associated
 with $X_{2}$-Laguerre Polynomials}
\author{Toshiaki Tanaka}
\email{ttanaka@mail.ncku.edu.tw}
\affiliation{Department of Physics, National Cheng Kung University,\\
 Tainan 701, Taiwan, R.O.C.\\
 National Center for Theoretical Sciences, Taiwan, R.O.C.}


\begin{abstract}

We construct a new family of quasi-solvable and $\cN$-fold
supersymmetric quantum systems where each Hamiltonian preserves
an exceptional polynomial subspace of codimension $2$. We show
that the family includes as a particular case the recently
reported rational radial oscillator potential whose eigenfunctions
are expressed in terms of the $X_{2}$-Laguerre polynomials of
the second kind. In addition, we find that the two kinds of
the $X_{2}$-Laguerre polynomials are ingeniously connected with
each other by the $\cN$-fold supercharge.

\end{abstract}






\maketitle

\section{Introduction}
\label{sec:intro}

Recently, a new type of polynomial systems called \emph{exceptional}
polynomials has attracted attention in mathematical physics. Roughly
speaking, they are an infinite sequence of polynomials which are
eigenfunctions of a second-order differential operator but which
have a non-zero lowest degree $m>0$. They are usually generated
by an infinite sequence of exceptional polynomial subspaces of
a fixed codimension $m$. According to the definition in
Refs.~\cite{GKM07,GKM08a}, a $k$-dimensional subspace $M_{k}$ of
the $\cN$-dimensional type A monomial subspace $\tcV_{\cN}^{(\rmA)}$:
\begin{align}
M_{k}\subset\tcV_{\cN}^{(\rmA)}[z]=\braket{1,z,\dots,z^{\cN-1}},
\end{align}
is called an \emph{exceptional} polynomial subspace of
codimension $m=\cN-k$ or an $X_{m}$ subspace if the linear
space $\cD_{2}(M_{k})$ of second-order linear differential
operators which preserve $M_{k}$ is \emph{not} a subspace of
the linear space $\cD_{2}(\tcV_{\cN}^{(\rmA)})$ of those which
preserve $\tcV_{\cN}^{(\rmA)}$, namely, $\cD_{2}(M_{k})
\not\subseteq\cD_{2}(\tcV_{\cN}^{(\rmA)})$. Various mathematical
results on the $X_{1}$ polynomials and $X_{1}$ subspaces are
explored in Refs.~\cite{GKM07,GKM08a,GKM08b}.

On the other hand, some exactly solvable quantum mechanical
potentials whose eigenfunctions are expressed in terms of
$X_{m}$ polynomials have been reported in the last few
years~\cite{Qu08b,Qu09,OS09}. All of them were constructed
using the technique of supersymmetric (SUSY) quantum
mechanics~\cite{Wi81,CKS95,Ju96,Ba00}
and have shape invariance~\cite{Ge83}. Those findings imply
wide applicability of such mathematical concepts in various
physics and mathematical sciences. Thus, it would be interesting
to see whether there exist other exactly solvable and/or less
restrictive quasi-exactly solvable models \cite{TU87,Us94}
associated with $X_{m}$ polynomials and $X_{m}$ subspaces.

In this respect, the recent research developments have shown that
the framework of $\cN$-fold SUSY~\cite{AIS93,AST01b,AS03,GT05}
is remarkably useful for investigating such issues (for a review
of $\cN$-fold SUSY see Ref.~\cite{Ta09}). This originates from
the important fact that $\cN$-fold SUSY is essentially equivalent
to \emph{weak quasi-solvability} (for the precise definitions of
the hierarchy of solvability, see, e.g., Refs.~\cite{Ta06a,Ta09})
proved in Ref.~\cite{AST01b} (for the equivalence in specific
models, see also references cited therein). Since weak
quasi-solvability includes solvability, on the one hand, and
shape invariance is a sufficient condition for
solvability\footnote{It should be noted that shape invariance
is \emph{not} a sufficient condition for \emph{exact solvability};
shape invariance, as well as solvability, is a local concept
while exact solvability is a global one. For the importance
of recognizing the difference between local and global concepts,
see, e.g., Ref.~\cite{GT06}.}, on the other hand, shape invariance
always implies $\cN$-fold SUSY,
\[
(\text{$\cN$-fold SUSY})\equiv(\text{Weak quasi-solvability})
 \supset(\text{Solvability})\supset(\text{Shape invariance}).
\]
Hence, all the shape invariant potentials reported in
Refs.~\cite{Qu08b,Qu09,OS09} must have $\cN$-fold SUSY.
The ones whose eigenfunctions are expressed in terms of
the $X_{1}$-Laguerre or $X_{1}$-Jacobi polynomials are evidently
preserve exceptional polynomial subspaces of codimension $1$.
On the other hand, according to Theorem~1.4 in Ref.~\cite{GKM08a},
there is only one exceptional polynomial subspace of codimension
$1$ up to the projective equivalence and its representative can be
chosen as type B monomial space $\tcV_{\cN}^{(\textrm{B})}$,
\begin{align}
\tcV_{\cN}^{(\textrm{B})}[z]=\braket{1,z,\dots,z^{\cN-2},z^{\cN}}.
\end{align}
This monomial space is first considered in Ref.~\cite{PT95} in the
context of the classification of monomial spaces preserved by
second-order linear ordinary differential operators.
It was shown that any Hamiltonian which preserves the type B
monomial space belongs to type B $\cN$-fold SUSY~\cite{GT04,GT05}.
Thus, we come to the conclusion that all of the models whose
eigenfunctions are expressed in terms of the $X_{1}$-Laguerre or
$X_{1}$-Jacobi polynomials in Refs.~\cite{Qu08b,Qu09} belong
to type B $\cN$-fold SUSY up to the projective equivalence.
In fact, the rational potential $V(x)$ in Ref.~\cite{Qu08b},
Eq.~(8), whose eigenfunctions are written in terms of the
$X_{1}$-Laguerre polynomials coincides with one
of the type B $\cN$-fold SUSY models $V_{\cN}^{+}(q)$ in
Ref.~\cite{GT04}, Eq.~(4.2). Explicitly, they are identical to
each other up to a multiplicative factor $V(x)=2V_{\cN}^{+}(x)$
with the following parameter relations:
\begin{align}
b_{1}=\omega,\qquad h_{0}=\frac{2l+1}{\omega},\qquad
 R=\frac{\omega}{\cN}-\frac{\omega(2\cN-2l-1)}{4}.
\end{align}
Hence, the framework of $\cN$-fold SUSY would provide a powerful
tool for investigating and constructing solvable and quasi-solvable
quantum systems associated not only with monomial spaces but also
with exceptional polynomial subspaces.
On the other hand, it has not been reported yet, to the best of
our knowledge, any $\cN$-fold SUSY where a component Hamiltonian
preserves any $X_{m}$ subspace of codimension $m>1$.

In this paper, we construct for the first time a family of
quasi-solvable and $\cN$-fold SUSY quantum systems where each
Hamiltonian preserves an exceptional polynomial subspace of
codimension $2$. We rely on the algorithmic construction developed
in Ref.~\cite{GT05}. The resulting $\cN$-fold SUSY systems turn to
include as a particular case the rational shape invariant potential
whose eigenfunctions are expressed in terms of the $X_{2}$-Laguerre
polynomials of the second kind in Ref.~\cite{Qu09}. Furthermore,
we find, in particular, that the two $X_{2}$ subspaces connected by
the $\cN$-fold supercharge would generate the two different kinds
of the $X_{2}$-Laguerre polynomials found in the latter reference.

We organize the paper as follows. In Section~\ref{sec:X2sub}, we
introduce a finite-dimensional polynomial space which turns out to
be an exceptional polynomial subspace of codimension $2$. Then, we
present quasi-solvable operators which leave the latter space
invariant. With the set of the polynomial space and the
quasi-solvable operators, we construct in Section~\ref{sec:CNfSS}
a new type of $\cN$-fold SUSY systems by applying the algorithm
developed in Ref.~\cite{GT05}. We present the pair of $\cN$-fold
SUSY Hamiltonians and the $\cN$-fold supercharges in closed form.
As a by-product, we obtain another exceptional polynomial subspace
of codimension $2$ and a set of quasi-solvable operators which
preserve it. In Section~\ref{sec:QSpot}, we exhibit a couple of
examples of quantum mechanical systems which possess the new
$\cN$-fold SUSY. In particular, we show that one of them coincides
with the rational potential model whose eigenfunctions are
expressed in terms of the $X_{2}$-Laguerre polynomials
of the second kind recently found in Ref.~\cite{Qu09}. Finally,
we summarize the results and discuss various perspectives of
future issues in Section~\ref{sec:discus}.

\section{An $X_{2}$ Polynomial Subspace and Its Quasi-solvable
 Operators}
\label{sec:X2sub}

Our starting point is to consider a vector space $\tcV_{\cN}^{-}$
of finite dimension $\cN$,
\begin{align}
\tcV_{\cN}^{-}=\tcV_{\cN}^{(\rmX_{2a})}[z;\alpha]=
 \braket{\tvph_{1}(z;\alpha),\dots,\tvph_{\cN}(z;\alpha)},
\label{eq:X2vec}
\end{align}
spanned by polynomials $\tvph_{n}(z;\alpha)$ of degree $n+1$ in $z$
as follows:
\begin{align}
\tvph_{n}(z;\alpha)=(\alpha+n-2)z^{n+1}+2(\alpha+n-1)(\alpha-1)z^{n}
 +(\alpha+n)(\alpha-1)\alpha z^{n-1},
\label{eq:X2pol}
\end{align}
where $\alpha(\neq0,1)$ is a parameter. When $\alpha=1$, then
$\tvph_{n}(z;1)=(n-1)z^{n+1}$ by definition. Hence, the vector space
$\tcV_{\cN}^{(\rmX_{2a})}[z;\alpha]$ in (\ref{eq:X2vec}) reduces to
a space which is equivalent to type A monomial space
$\tcV_{\cN-1}^{(\rmA)}$ (cf., Ref.~\cite{GT05}),
\begin{align}
\tcV_{\cN}^{(\rmX_{2a})}[z;1]=z^{3}\braket{1,z,\dots,z^{\cN-2}}
 =z^{3}\,\tcV_{\cN-1}^{(\rmA)}[z].
\end{align}
Similarly, when $\alpha=0$, then $\tvph_{n}(z;0)=(n-2)z^{n+1}-2(n-1)
z^{n}$ by definition and, in particular, $2\tvph_{1}(z;0)=\tvph_{2}(z;0)
=-2z^{2}$. On the other hand, we can inductively prove the following
formula:
\begin{align}
(n-1)\sum_{k=3}^{n}\frac{2^{n-k}}{(k-1)(k-2)}\tvph_{k}(z;0)=z^{n+1}
 -(n-1)2^{n-2}z^{3}.
\end{align}
Hence, in the case of $\alpha=0$ we have
\begin{align}
\tcV_{\cN}^{(\rmX_{2a})}[z;0]=z^{2}\,V_{\cN-1}
 =z^{2}\braket{1,z^{2}-4z,\ldots,z^{\cN-1}-(\cN-1)2^{\cN-2}z}.
\end{align}
By Proposition 2.5 in Ref.~\cite{GKM08a} with $n=\cN-1$, $\lambda=1$,
$\mu=0$, and $\beta_{j}=-j\,2^{j-1}$, the $(\cN-1)$-dimensional
linear space $V_{\cN-1}\subset\tcV_{\cN}^{(\rmA)}[z]$ introduced above
is a polynomial subspace of codimension $1$ and its fundamental
covariant $q_{V_{\cN-1}}(z)$ is given by
\begin{align}
q_{V_{\cN-1}}(z)=-(\cN-1)(z-2)^{\cN-2}.
\end{align}
That is, it has a root of multiplicity $\cN-2=n-1$ at $z=2$ and thus
the space $V_{\cN-1}$ is an exceptional polynomial subspace of
codimension $1$ which is projectively equivalent to type B
monomial space~\cite{GKM08a}
\begin{align}
V_{\cN-1}\sim\tcV_{\cN-1}^{(\mathrm{B})}[z]
 =\braket{1,z,\dots,z^{\cN-3},z^{\cN-1}}.
\end{align}
Therefore, the constraint $\alpha\neq0,1$ prevents the space
$\tcV_{\cN}^{(\rmX_{2a})}[z;\alpha]$ from reducing to the well-studied
monomial spaces of types A and B (for the latter issues, see
Refs.~\cite{GKO93,PT95,Ta03a,GT04,GKM07,GKM08a}).
In addition, we can assume $\cN>2$ without any loss of generality.
Indeed, when $\cN=1,2$, the space $\tcV_{\cN}^{(\rmX_{2a})}[z;\alpha]$
is also essentially equivalent to type A monomial space as
\begin{align}
\tcV_{1}^{(\rmX_{2a})}[z;\alpha]=\tvph_{1}(z;\alpha)
 \tcV_{1}^{(\rmA)}[z],\qquad\tcV_{2}^{(\rmX_{2a})}[z;\alpha]
 =\tvph_{1}(z;\alpha)\tcV_{2}^{(\rmA)}[w],
\end{align}
where $w=\tvph_{2}(z;\alpha)/\tvph_{1}(z;\alpha)$. Thus, we hereafter
assume $\cN>2$.

We shall look for a vector space of linear differential operators
of (at most) second order which preserve the polynomial space
$\tcV_{\cN}^{-}$. One of the characteristic features of the set of
polynomials $\tvph_{n}(z;\alpha)$ defined in (\ref{eq:X2pol}) is
the following factorization under the action of the operator
$\del_{z}-1$:
\begin{align}
(\del_{z}-1)\tvph_{n}(z;\alpha)=-[(\alpha+n-2)z
 -(n-1)(\alpha+n)]z^{n-2}f(z;\alpha),
\label{eq:fact1}
\end{align}
where the common factor $f(z;\alpha)$ is an $n$-independent polynomial
of second degree in $z$ given by
\begin{align}
f(z;\alpha)=z^{2}+2(\alpha-1)z+(\alpha-1)\alpha.
\label{eq:deff}
\end{align}
Taking into account the factorization property (\ref{eq:fact1}), we
easily see that any (at most) second-order linear differential operator
$J$ of the following form:
\begin{align}
J=p_{4}(z)\del_{z}^{\,2}+p_{3}(z)\del_{z}+p_{2}(z)
 +\frac{p_{1}(z)}{f(z;\alpha)}\left(\del_{z}-1\right),
\label{eq:opf}
\end{align}
where $p_{i}(z)$ ($i=1,\dots,4$) are all polynomials in $z$, maps all
the polynomials $\tvph_{n}(z;\alpha)$ ($n=1,2,3,\ldots$) to other
polynomials which are not necessarily elements of $\tcV_{\cN}^{-}$.
The latter fact is rather considered as a necessary condition for
an operator to preserve the vector space $\tcV_{\cN}^{-}$. Hence,
we restrict ourselves to considering linear operators of the form
(\ref{eq:opf}). As we will show shortly, quasi-solvable operators
of the form (\ref{eq:opf}) which preserve $\tcV_{\cN}^{-}=
\tcV_{\cN}^{(\rmX_{2a})}[z;\alpha]\subset\tcV_{\cN+2}^{(\rmA)}[z]$
consist of those which have non-trivial $p_{1}(z)$, and thus do not
preserve type A monomial space $\tcV_{\cN+2}^{(\rmA)}[z]$,
namely, $\cD_{2}(\tcV_{\cN}^{(\rmX_{2a})})\not\subseteq\cD_{2}(
\tcV_{\cN+2}^{(\rmA)})$. 
Hence, the polynomial subspace $\tcV_{\cN}^{(\rmX_{2a})}[z;\alpha]$
belongs to an exceptional polynomial subspace of codimension $2$
which is denoted by $X_{2}$ in Ref.~\cite{GKM08a}.
For $\cN>2$, we find that there are four linearly independent such
quasi-solvable operators which leave the space $\tcV_{\cN}^{-}$
invariant. The first operator denoted by $J_{1}$ is given by
\begin{align}
J_{1}=z\del_{z}^{\,2}-(z-\alpha+3)\del_{z}
 +\frac{4(\alpha-1)(z+\alpha)}{f(z;\alpha)}\left(\del_{z}-1\right).
\end{align}
The action of $J_{1}$ on $\tvph_{n}(z;\alpha)$ ($n=1,2,3,\ldots$)
reads as
\begin{align}
J_{1}\tvph_{n}(z;\alpha)=-(n+1)\tvph_{n}(z;\alpha)
 +(n-1)(\alpha+n)\tvph_{n-1}(z;\alpha).
\end{align}
Hence, it preserves not only $\tcV_{\cN}^{-}$ for a specific value of
$\cN$ but also an infinite flag of the spaces
\begin{align}
\tcV_{1}^{-}\subset\tcV_{2}^{-}\subset\cdots\subset\tcV_{\cN}^{-}
 \subset\cdots.
\end{align}
Therefore, the operator $J_{1}$ is not only quasi-solvable but also
solvable\footnote{This significant characterization of solvability
first appeared in Ref.~\cite{Tu94} but unfortunately without
sufficient appreciation of the difference between solvability and
exact solvability.}. The second operator, denoted by $J_{2}$, which
preserves $\tcV_{\cN}^{-}$ is given by
\begin{align}
J_{2}=&\,[z^{2}+(\alpha-1)(\alpha+\cN-1)]\del_{z}^{\,2}-[z^{2}+(\cN+1)z
 +(\alpha-1)(3\alpha+3\cN-7)]\del_{z}\notag\\
&+(\cN+1)z-4(\alpha-1)\frac{(2\alpha+\cN-3)z+(\alpha-1)(2\alpha+\cN-1)}{
 f(z;\alpha)}\left(\del_{z}-1\right).
\end{align}
The action of $J_{2}$ on $\tvph_{n}(z;\alpha)$ ($n=1,2,3,\ldots$) reads as
\begin{align}
\lefteqn{
(\alpha+n-2)(\alpha+n-1)J_{2}\tvph_{n}(z;\alpha)=-(n-\cN)
 (\alpha+n-2)^{2}\tvph_{n+1}(z;\alpha)+s^{-}(n,\alpha,\cN)
 }\hspace{30pt}\notag\\
&\times\tvph_{n}(z;\alpha)-(n-1)(\alpha-1)(\alpha+\cN-1)
 [3\alpha^{2}+6(n-1)\alpha+3n^{2}-6n+4]\tvph_{n-1}(z;\alpha)\notag\\
&+(n-1)(n-2)(\alpha-1)(\alpha+\cN-1)(\alpha+n-1)(\alpha+n)
 \tvph_{n-2}(z;\alpha),
\end{align}
where $s^{-}(n,\alpha,\cN)$ is given by
\begin{align}
s^{-}(n,\alpha,\cN)=&\,2\alpha^{3}+[n^{2}-(\cN-4)n-\cN-7]\alpha^{2}
 \notag\\
&+[2n^{3}-(2\cN+1)n^{2}+(\cN-12)n+5\cN+9]\alpha\notag\\
&+n^{4}-(\cN+3)n^{3}+(2\cN-1)n^{2}+(\cN+9)n-4(\cN+1).
\end{align}
Hence, it certainly satisfies $J_{2}\tcV_{\cN}^{-}\subset\tcV_{\cN}^{-}$
only for a specific value of $\cN\in\bbN$. The third operator, denoted by
$J_{3}$, leaving $\tcV_{\cN}^{-}$ invariant is given by
\begin{align}
J_{3}=&\,(z+2\alpha+\cN-1)z^{2}\del_{z}^{\,2}
 +\bigl\{(\alpha-\cN-2)z^{2}+[3\alpha^{2}+(\cN-2)\alpha-2(\cN+1)]z
 \notag\\[5pt]
&+4(\alpha-1)(\alpha+\cN+1)\bigr\}\del_{z}-(\cN+1)(\alpha-2)z\notag\\
&-4(\alpha-1)\frac{(\alpha^{2}+\cN\alpha-2\cN-2)z
 +\alpha(\alpha-1)(\alpha+\cN+1)}{f(z;\alpha)}\left(\del_{z}-1\right).
\end{align}
The action of $J_{3}$ on $\tvph_{n}(z;\alpha)$ ($n=1,2,3,\ldots$)
admits a relatively simple form as
\begin{align}
J_{3}\tvph_{n}(z;\alpha)=&\,(n-\cN)(\alpha+n-2)\tvph_{n+1}(z;\alpha)
 +\bigl\{(3n+1)\alpha^{2}+[2n^{2}+(\cN-4)n\notag\\
&+3\cN+2]\alpha+(\cN-1)n(n-1)-4(\cN+1)\bigr\}\tvph_{n}(z;\alpha),
\end{align}
and thus it also satisfies $J_{3}\tcV_{\cN}^{-}\subset\tcV_{\cN}^{-}$
only for a specific value of $\cN\in\bbN$. The fourth quasi-solvable
operator denoted by $J_{4}$ has the most complicated form as follows:
\begin{align}
\lefteqn{(2\alpha+\cN-1)
J_{4}=[(2\alpha+\cN-1)z+3\alpha^{2}+(3\cN-2)\alpha
 +\cN(\cN-1)]z^{3}\del_{z}^{\,2}}\hspace{30pt}\notag\\[5pt]
&-B_{4}^{-}(z;\alpha,\cN)\del_{z}+\cN(\cN+1)(2\alpha+\cN-1)z^{2}
 +(\cN+1)(\alpha-1)[3\alpha^{2}\notag\\
&+2(3\cN-8)\alpha+2(\cN-4)(\cN-1)]z
 +4(\alpha-1)^{2}\frac{D_{4}^{-}(z;\alpha,\cN)}{f(z;\alpha)}
 \left(\del_{z}-1\right),
\end{align}
where $B_{4}^{-}(z;\alpha,\cN)$ and $D_{4}^{-}(z;\alpha,\cN)$ are,
respectively, given by
\begin{align}
\lefteqn{
B_{4}^{-}(z;\alpha,\cN)=2\cN(2\alpha+\cN-1)z^{3}+[3\alpha^{3}+(9\cN-19)
 \alpha^{2}+(5\cN^{\,2}-18\cN+24)\alpha}\hspace{30pt}\notag\\
&+(\cN-1)(\cN^{\,2}-2\cN+8)]z^{2}+(\alpha-1)[7\alpha^{3}+(14\cN-15)
 \alpha^{2}\notag\\
&+(3\cN+1)(3\cN-8)\alpha+2(\cN-4)(\cN^{\,2}-1)]z+(\alpha-1)[2\alpha^{4}
 +(5\cN+1)\alpha^{3}\notag\\
&+(4\cN^{\,2}+3\cN+13)\alpha^{2}+(\cN^{\,3}+2\cN^{\,2}-7\cN-36)
 \alpha-8(\cN-1)(\cN+2)]
\end{align}
and
\begin{align}
D_{4}^{-}(z;\alpha,\cN)=&\,[\alpha^{3}+(2\cN+3)\alpha^{2}+(\cN^{\,2}
 -5\cN-16)\alpha-4(\cN-1)(\cN+2)]z\notag\\
&+\alpha[\alpha^{3}+(2\cN+3)\alpha^{2}+(\cN^{\,2}
 -2\cN-9)\alpha-2(\cN-1)(\cN+2)].
\end{align}
The action of $J_{4}$ on $\tvph_{n}(z;\alpha)$ ($n=1,2,3,\ldots$) reads as
\begin{align}
\lefteqn{
(\alpha+n-1)(\alpha+n)J_{4}\tvph_{n}(z;\alpha)=(n-\cN)(n-\cN+1)
 (\alpha+n-2)(\alpha+n-1)\tvph_{n+2}(z;\alpha)
}\hspace{110pt}\notag\\
&-(n-\cN)t_{1}^{-}(n,\alpha,\cN)\tvph_{n+1}(z;\alpha)-(\alpha-1)
 t_{2}^{-}(n,\alpha,\cN)\tvph_{n}(z;\alpha)\notag\\
&-(n-1)\alpha(\alpha-1)(\alpha+\cN-1)(\alpha+\cN)(\alpha+n)^{2}
 \tvph_{n-1}(z;\alpha),
\end{align}
and thus it surely preserves the space $\tcV_{\cN}^{-}$ for a specific
value of $\cN\in\bbN$. In the above, $t_{1}^{-}(n,\alpha,\cN)$ and
$t_{2}^{-}(n,\alpha,\cN)$ are, respectively, given by
\begin{align}
\lefteqn{
(2\alpha+\cN-1)t_{1}^{-}(n,\alpha,\cN)=3\alpha^{5}+(3n+6\cN-20)
 \alpha^{4}-[3n^{2}-(9\cN-26)n}\hspace{30pt}\notag\\
&-2\cN^{\,2}+27\cN-50]\alpha^{3}-[3n^{3}+4n^{2}-(3\cN^{\,2}
 -35\cN+52)n+7\cN^{\,2}-49\cN+52]\alpha^{2}\notag\\
&-[(3\cN-2)n^{3}+(7\cN-10)n^{2}+2(4\cN^{\,2}-21\cN+18)n-2(\cN-1)
 (5\cN-12)]\alpha\notag\\
&-(\cN-1)(n-1)[\cN n^{2}+2(\cN-2)n-4(\cN-1)]
\end{align}
and
\begin{align}
\lefteqn{
(2\alpha+\cN-1)t_{2}^{-}(n,\alpha,\cN)=(7n+1)\alpha^{5}+2[7n^{2}
 +(7\cN-8)n+2\cN+6]\alpha^{4}}\hspace{30pt}\notag\\
&+[7n^{3}+28(\cN-1)n^{2}+(9\cN^{\,2}-19\cN+48)n
 +3\cN^{\,2}-19\cN-49]\alpha^{3}\notag\\
&+[(14\cN-11)n^{3}+18(\cN^{\,2}-2\cN+2)n^{2}+(2\cN^{\,3}-7\cN^{\,2}
 +\cN-97)n\notag\\
&-19\cN^{\,2}+7\cN+52]\alpha^{2}+[(\cN-1)(9\cN-4)n^{3}
 +2(2\cN-7)(\cN^{\,2}+3)n^{2}\notag\\
&\quad-(2\cN^{\,3}+15\cN^{\,2}+13\cN-70)n-2(\cN-1)(\cN^{\,2}-5\cN-8)]
 \alpha\notag\\
&+2(\cN-1)n(n-1)[\cN(\cN-1)n-4(\cN+2)].
\end{align}
Hence, the most general quasi-solvable operator $\tH^{-}$ of the form
(\ref{eq:opf}) which preserves the space $\tcV_{\cN}^{-}$ ($\cN>2$) is
given by
\begin{align}
\tH^{-}&=-\sum_{i=1}^{4}a_{i}J_{i}-c_{0}
 \notag\\
&=-A(z)\frac{\rmd^{2}}{\rmd z^{2}}-\left[\tB(z)+\frac{4(\alpha-1)D(z)
 }{f(z;\alpha)}\right]\frac{\rmd}{\rmd z}-\tC(z)+\frac{4(\alpha-1)D(z)
 }{f(z;\alpha)},
\label{eq:tH-}
\end{align}
where $a_{i}$ ($i=1,\dots,4$) and $c_{0}$ are constants,
while $A(z)$, $\tB(z)$, $\tC(z)$, and $D(z)$ are polynomials in $z$
given by
\begin{align}
A(z)=&\,a_{4}z^{4}+\left[\frac{3\alpha^{2}+(3\cN-2)\alpha+\cN(\cN-1)}{
 2\alpha+\cN-1}a_{4}+a_{3}\right]z^{3}+[(2\alpha+\cN-1)a_{3}+a_{2}]z^{2}
 \notag\\
&+a_{1}z+(\alpha-1)(\alpha+\cN-1)a_{2},
\label{eq:defAz}\\
\tB(z)=&-\frac{B_{4}^{-}(z;\alpha,\cN)}{2\alpha+\cN-1}a_{4}
 +[(\alpha-\cN-2)a_{3}-a_{2}]z^{2}\notag\\
&+\bigl\{[3\alpha^{2}+(\cN-2)\alpha-2(\cN+1)]a_{3}-(\cN+1)a_{2}
 -a_{1}\bigr\}z\notag\\[5pt]
&+4(\alpha-1)(\alpha+\cN+1)a_{3}-(\alpha-1)(3\alpha+3\cN-7)a_{2}
 +(\alpha-3)a_{1},\\[5pt]
\tC(z)=&\,\cN(\cN+1)a_{4}z^{2}+(\cN+1)\biggl[(\alpha-1)
 \frac{3\alpha^{2}+2(3\cN-8)\alpha+2(\cN-4)(\cN-1)}{2\alpha+\cN-1}
 a_{4}\notag\\
&-(\alpha-2)a_{3}+a_{2}\biggr]z+c_{0},\\
D(z)=&\,\frac{(\alpha-1)D_{4}^{-}(z;\alpha,\cN)}{2\alpha+\cN-1}a_{4}
 -[(\alpha^{2}+\cN\alpha-2\cN-2)a_{3}+(2\alpha+\cN-3)a_{2}-a_{1}]z
 \notag\\
&-\alpha(\alpha-1)(\alpha+\cN+1)a_{3}-(\alpha-1)(2\alpha+\cN-1)a_{2}
 +\alpha a_{1}.
\label{eq:defDz}
\end{align}
The operator $\tH$ becomes solvable only when
\begin{align}
a_{4}=a_{3}=a_{2}=0.
\label{eq:scond}
\end{align}

\section{Construction of $\cN$-fold Supersymmetric Systems}
\label{sec:CNfSS}

Now that we have constructed the set of quasi-solvable operators
which leave the space $\tcV_{\cN}^{-}$ invariant, we are in a position
to construct $\cN$-fold supersymmetric systems associated with
the latter space by applying the systematic algorithm developed
in Ref.~\cite{GT05}. The first step is to construct an $\cN$th-order
linear differential operator $\tP_{\cN}^{-}$ which annihilate
the space $\tcV_{\cN}^{-}$ and which has the following form (cf.,
Eq.~(2.29) in Ref.~\cite{GT05}):
\begin{align}
\tP_{\cN}^{-}&=z'(q)^{\cN}\left(\frac{\rmd^{\cN}}{\rmd z^{\cN}}
 +\sum_{k=0}^{\cN-1}\tw_{k}^{[\cN]}(z)\frac{\rmd^{k}}{\rmd z^{k}}
 \right).
\label{eq:tPN-}
\end{align}
where $z(q)$ denotes the change of variable connecting the variable
$z$ with a physical coordinate $q$, and $z'(q)$ is the first
derivative of $z(q)$ with respect to $q$. To construct the operator
$\tP_{\cN}^{-}$, we first note that the following first-order
linear differential operator $\tA^{-}(\alpha)$ plays a role of
lowering operator for all the polynomials $\tvph_{n}(z;\alpha)$,
\begin{align}
\tA^{-}(\alpha)=\frac{f(z;\alpha+1)}{f(z;\alpha)}\left(
 \frac{\rmd}{\rmd z}-\frac{f'(z;\alpha+1)}{f(z;\alpha+1)}\right),
\label{eq:defA-}
\end{align}
where $f(z;\alpha)$ is the polynomial introduced in (\ref{eq:deff})
and the prime denotes derivative with respect to $z$.
In fact, the action of $\tA^{-}(\alpha)$ on $\tvph_{n}(z;\alpha)$
for an arbitrary $n=1,2,3,\ldots$ reads as
\begin{align}
\tA^{-}(\alpha)\tvph_{n}(z;\alpha)=(n-1)\tvph_{n-1}(z;\alpha+1).
\end{align}
Using the latter formula, we easily prove the following formula by
induction:
\begin{align}
\left(\prod_{k=0}^{m-1}\tA^{-}(\alpha+k)\right)\tvph_{n}(z;\alpha)
 =\frac{\Gamma(n)}{\Gamma(n-m)}\tvph_{n-m}(z;\alpha+m),
\end{align}
where $\Gamma$ denotes the gamma function and the product of operators
is defined by
\begin{align}
\prod_{k=k_{0}}^{k_{1}}A_{k}\equiv A_{k_{1}}A_{k_{1}-1}\dots A_{k_{0}}.
\end{align}
Then, we immediately know that the properly normalized operator
$\tP_{\cN}^{-}$ of the form (\ref{eq:tPN-}) whose kernel is the
polynomial space $\tcV_{\cN}^{-}$ is given by
\begin{align}
\tP_{\cN}^{-}&=z'(q)^{\cN}\frac{f(z;\alpha)}{f(z;\alpha+\cN)}
 \prod_{k=0}^{\cN-1}\tA^{-}(\alpha+k)\notag\\
&=z'(q)^{\cN}\frac{f(z;\alpha)}{f(z;\alpha+\cN)}\prod_{k=0}^{\cN-1}
 \frac{f(z;\alpha+k+1)}{f(z;\alpha+k)}\left(\frac{\rmd}{\rmd z}
 -\frac{f'(z;\alpha+k+1)}{f(z;\alpha+k+1)}\right).
\label{eq:tPN-2}
\end{align}
The next task is to calculate the coefficient $\tw_{\cN-1}^{[\cN]}(z)$
of the $(\cN-1)$th-order differential operator $\del_{z}^{\,\cN-1}$ in
$\tP_{\cN}^{-}$ defined by (\ref{eq:tPN-}). For the latter purpose, we
first derive a formula for the calculation of $\tw_{\cN-1}^{[\cN]}(z)$
in a general setting,
\begin{align}
\prod_{k=0}^{\cN-1}f_{k}(z)\left(\frac{\rmd}{\rmd z}+g_{k}(z)\right)
 =\left(\prod_{k=0}^{\cN-1}f_{k}(z)\right)\left(
 \frac{\rmd^{\cN}}{\rmd z^{\cN}}+\tw_{\cN-1}^{[\cN]}(z)
 \frac{\rmd^{\cN-1}}{\rmd z^{\cN-1}}+\cdots\right).
\end{align}
It is an easy task to show inductively with respect to $\cN$
the following recursion relation:
\begin{align}
\tw_{\cN}^{[\cN+1]}(z)=\tw_{\cN-1}^{[\cN]}(z)+g_{\cN}(z)
 +\sum_{k=0}^{\cN-1}\frac{f'_{k}(z)}{f_{k}(z)},\qquad
\tw_{0}^{[1]}(z)=g_{0}(z).
\end{align}
Its general solution is given by
\begin{align}
\tw_{\cN-1}^{[\cN]}(z)=\sum_{k=0}^{\cN-1}\left[g_{k}(z)+(\cN-1-k)
 \frac{f'_{k}(z)}{f_{k}(z)}\right].
\label{eq:wN-1}
\end{align}
In our present case, we read from (\ref{eq:defA-}) and (\ref{eq:tPN-2})
that the functions $f_{k}(z)$ and $g_{k}(z)$ are given by
\begin{align}
f_{k}(z)=\frac{f(z;\alpha+k+1)}{f(z;\alpha+k)},\qquad
 g_{k}(z)=-\frac{f'(z;\alpha+k+1)}{f(z;\alpha+k+1)}.
\label{eq:fkgk}
\end{align}
Then, for the latter $f_{k}(z)$ and $g_{k}(z)$, we have
\begin{align}
\sum_{k=0}^{\cN-1}\frac{f'_{k}(z)}{f_{k}(z)}=\frac{\rmd}{\rmd z}\ln
 \left(\prod_{k=0}^{\cN-1}\frac{f(z;\alpha+k+1)}{f(z;\alpha+k)}\right)
 =\frac{f'(z;\alpha+\cN)}{f(z;\alpha+\cN)}
 -\frac{f'(z;\alpha)}{f(z;\alpha)}
\end{align}
and
\begin{align}
\sum_{k=0}^{\cN-1}k\frac{f'_{k}(z)}{f_{k}(z)}&=\frac{\rmd}{\rmd z}\ln
 \left(\prod_{k=0}^{\cN-1}\frac{f(z;\alpha+k+1)^{k}}{f(z;\alpha+k)^{k}}
 \right)\notag\\
&=(\cN-1)\frac{f'(z;\alpha+\cN)}{f(z;\alpha+\cN)}
 -\sum_{k=1}^{\cN-1}\frac{f'(z;\alpha+k)}{f(z;\alpha+k)}.
\label{eq:sum2}
\end{align}
Substituting (\ref{eq:fkgk})--(\ref{eq:sum2}) into (\ref{eq:wN-1}),
we obtain
\begin{align}
\tw_{\cN-1}^{[\cN]}(z)=-(\cN-1)\frac{f'(z;\alpha)}{f(z;\alpha)}
 -\frac{f'(z;\alpha+\cN)}{f(z;\alpha+\cN)}.
\label{eq:wN-1a}
\end{align}
With the obtained function $\tw_{\cN-1}^{[\cN]}(z)$, the $\cN$-fold SUSY
pair of gauged Hamiltonians $\tH^{-}$ and $\bar{H}^{+}$ are expressed as
(cf., Eq.~(2.45) in Ref.~\cite{GT05})
\begin{align}
\bar{\tH}^{\pm}=&-A(z)\frac{\rmd^{2}}{\rmd z^{2}}+\left(\frac{\cN-2}{2}
 A'(z)\pm Q(z)\right)\frac{\rmd}{\rmd z}-C(z)\notag\\
&-(1\pm 1)\left(\frac{\cN-1}{2}Q'(z)
 -\frac{A'(z)\tw_{\cN-1}^{[\cN]}(z)}{2}
 -A(z)\tw_{\cN-1}^{[\cN]\prime}(z)\right),
\label{eq:btH+-}
\end{align}
where $Q(z)$ and $C(z)$ in the present case read as
\begin{align}
Q(z)&=\frac{\cN-2}{2}A'(z)+\tB(z)
 +\frac{4(\alpha-1)D(z)}{f(z;\alpha)},
\label{eq:defQz}\\
C(z)&=\tC(z)-\frac{4(\alpha-1)D(z)}{f(z;\alpha)}.
\label{eq:defCz}
\end{align}
It is worth studying another linear space $\bar{\cV}_{\cN}^{+}$
preserved by the partner gauged Hamiltonian $\bar{H}^{+}$. The
latter space is characterized by the kernel of another $\cN$th-order
linear differential operator $\bar{P}_{\cN}^{+}$, namely, through
the relation $\ker\bar{P}_{\cN}^{+}=\bar{\cV}_{\cN}^{+}$. The operator
$\bar{P}_{\cN}^{+}$ is obtained from $\tP_{\cN}^{-}$ in (\ref{eq:tPN-})
by (cf., Eqs.~(2.30)--(2.32) in Ref.~\cite{GT05})
\begin{align}
\bar{P}_{\cN}^{+}&=(-1)^{\cN}z'(q)^{\cN-1}(\tP_{\cN}^{-})^{\rmT}
 z'(q)^{1-\cN}\notag\\
&=z'(q)^{\cN}\left(\frac{\rmd^{\cN}}{\rmd z^{\cN}}+\sum_{k=0}^{\cN-1}
 (-1)^{\cN-k}\frac{\rmd^{k}}{\rmd z^{k}}\tw_{k}^{[\cN]}(z)\right),
\end{align}
where the subscript $\rmT$ stands for the transposition in the physical
$q$-space. In the present case where $\tP_{\cN}^{-}$ is given by
(\ref{eq:tPN-2}), it reads as
\begin{align}
\bar{P}_{\cN}^{+}&=z'(q)^{\cN}\left[\prod_{k=0}^{\cN-1}\left(
 \frac{\rmd}{\rmd z}+\frac{f'(z;\alpha+\cN-k)}{f(z;\alpha+\cN-k)}\right)
 \frac{f(z;\alpha+\cN-k)}{f(z;\alpha+\cN-k-1)}\right]
 \frac{f(z;\alpha)}{f(z;\alpha+\cN)}\notag\\
&=z'(q)^{\cN}\frac{f(z;\alpha-1)}{f(z;\alpha)}\left(\prod_{k=0}^{\cN-1}
 \bar{A}^{+}(\alpha+\cN-k)\right)\frac{f(z;\alpha)}{f(z;\alpha+\cN-1)},
\label{eq:bPN+2}
\end{align}
where $\bar{A}^{+}(\alpha)$ is a first-order linear differential operator
defined by
\begin{align}
\bar{A}^{+}(\alpha)=\frac{f(z;\alpha-1)}{f(z;\alpha-2)}
 \left(\frac{\rmd}{\rmd z}+\frac{f'(z;\alpha)}{f(z;\alpha)}\right) .
\label{eq:defA+}
\end{align}
The vector space $\bar{\cV}_{\cN}^{+}$ annihilated by
$\bar{P}_{\cN}^{+}$ is obtained by integrating inductively
the differential equation $\bar{P}_{\cN}^{+}\bar{\chi}(z)=0$.
The result is
\begin{align}
f(z;\alpha)f(z;\alpha+\cN)\bar{\cV}_{\cN}^{+}=\tcV_{\cN}^{(\rmX_{2b})}
 [z;\alpha+\cN]=\braket{\bar{\chi}_{1}(z;\alpha+\cN),\dots,
 \bar{\chi}_{\cN}(z;\alpha+\cN)},
\label{eq:bVN+}
\end{align}
where each $\bar{\chi}_{n}(z;\alpha)$ is a polynomial of degree $n+1$
defined by
\begin{align}
\bar{\chi}_{n}(z;\alpha)=&\,(\alpha-n)(\alpha-n+1)z^{n+1}
 +2(\alpha-n-1)(\alpha-n+1)(\alpha-1)z^{n}\notag\\
&+(\alpha-n-1)(\alpha-n)(\alpha-1)\alpha z^{n-1}.
\label{eq:X2pol2}
\end{align}
We find that the obtained linear space $\bar{\cV}_{\cN}^{+}$ in
(\ref{eq:bVN+}) is such a space on which the operator $\bar{A}^{+}
(\alpha)$ introduced in (\ref{eq:defA+}) acts essentially as a
lowering operator. Indeed, we easily derive the following formula:
\begin{align}
\bar{A}^{+}(\alpha)\frac{\bar{\chi}_{n}(z;\alpha)}{f(z;\alpha-1)
 f(z;\alpha)}=(n-1)\frac{\bar{\chi}_{n-1}(z;\alpha-1)}{
 f(z;\alpha-2)f(z;\alpha-1)}.
\end{align}
With repeated applications of the latter formula, we obtain
\begin{multline}
\left(\prod_{k=0}^{m-1}\bar{A}^{+}(\alpha+\cN-k)\right)
 \frac{\bar{\chi}_{n}(z;\alpha+\cN)}{f(z;\alpha+\cN-1)f(z;\alpha+\cN)}
 =\\
\frac{\Gamma(n)}{\Gamma(n-m)}\frac{\bar{\chi}_{n-m}(z;\alpha+\cN-m)}{
 f(z;\alpha+\cN-m-1)f(z;\alpha+\cN-m)}.
\label{eq:ann+}
\end{multline}
Then, combining (\ref{eq:bPN+2}), (\ref{eq:bVN+}), and (\ref{eq:ann+}),
we eventually have
\begin{align}
\bar{P}_{\cN}^{+}\bar{\cV}_{\cN}^{+}\propto\left(\prod_{k=0}^{\cN-1}
 \bar{A}^{+}(\alpha+\cN-k)\right)\frac{\braket{\bar{\chi}_{1}
 (z;\alpha+\cN),\dots,\bar{\chi}_{\cN}(z;\alpha+\cN)}}{
 f(z;\alpha+\cN-1)f(z;\alpha+\cN)}=0.
\end{align}
As the gauged Hamiltonian $\bar{H}^{+}$ preserves the vector space
$\bar{\cV}_{\cN}^{+}$ defined by the relation in (\ref{eq:bVN+}),
namely, $\bar{H}^{+}\bar{\cV}_{\cN}^{+}\subset\bar{\cV}_{\cN}^{+}$,
it is evident that the linear operator
\begin{align}
\check{H}^{+}=f(z;\alpha)f(z;\alpha+\cN)\bar{H}^{+}f(z;\alpha+\cN)^{-1}
 f(z;\alpha)^{-1},
\label{eq:chH+}
\end{align}
preserves the polynomial subspace $\tcV_{\cN}^{(\rmX_{2b})}[z;\alpha
+\cN]\subset\tcV_{\cN+2}^{(\rmA)}[z]$, namely, $\check{H}^{+}
\tcV_{\cN}^{(\rmX_{2b})}[z;\alpha+\cN]\subset\tcV_{\cN}^{(\rmX_{2b})}
[z;\alpha+\cN]$. On the other hand, it immediately follows from the
form of $\bar{H}^{+}$ given by (\ref{eq:btH+-})--(\ref{eq:defCz}) that
the operator $\check{H}^{+}$ does not preserve the monomial space
$\tcV_{\cN+2}^{(\rmA)}$. In other words, $\cD_{2}(\tcV_{\cN}^{(\rmX_{2b})})
\not\subseteq\cD_{2}(\tcV_{\cN+2}^{(\rmA)})$. Hence, the linear space
$\tcV_{\cN}^{(\rmX_{2b})}$ spanned by the polynomials $\bar{\chi}_{n}$
in (\ref{eq:X2pol2}) provides another exceptional polynomial subspace
of codimension $2$. From (\ref{eq:btH+-}) and (\ref{eq:chH+}), the form
of the operator $\check{H}^{+}$ reads as
\begin{align}
\check{H}^{+}=-A(z)\frac{\rmd^{2}}{\rmd z^{2}}-B^{+}(z)
 \frac{\rmd}{\rmd z}-C^{+}(z),
\label{eq:chH+2}
\end{align}
where $B^{+}(z)$ and $C^{+}(z)$ are, respectively, given by
\begin{align}
B^{+}(z)=&-(\cN-2)A'(z)-\tB(z)-2\frac{A(z)f'(z;\alpha)+2(\alpha-1)D(z)
 }{f(z;\alpha)}\notag\\
&-2A(z)\frac{f'(z;\alpha+\cN)}{f(z;\alpha+\cN)}
\label{eq:defB+}
\end{align}
and
\begin{align}
C^{+}(z)=&\,(\cN-1)\left[\frac{\cN-2}{2}A''(z)+\tB'(z)\right]+\tC(z)
 +\frac{1}{f(z;\alpha)}\Bigl\{2(2\cN-3)A(z)\notag\\
&+\bigl[(2\cN-3)A'(z)+\tB(z)\bigr]f'(z;\alpha)-4(\alpha-1)\bigl[D(z)
 -(\cN-1)D'(z)\bigr]\Bigr\}\notag\\
&+\frac{2A(z)+[(\cN-1)A'(z)+\tB(z)]f'(z;\alpha+\cN)}{f(z;\alpha+\cN)}
 \notag\\
&-2\frac{A(z)f'(z;\alpha)+2(\alpha-1)D(z)}{f(z;\alpha)}\left[(\cN-2)
 \frac{f'(z;\alpha)}{f(z;\alpha)}-\frac{f'(z;\alpha+\cN)}{
 f(z;\alpha+\cN)}\right].
\label{eq:defC+}
\end{align}
In the above, the functions $A(z)$, $\tB(z)$, $\tC(z)$, and $D(z)$ are
given by (\ref{eq:defAz})--(\ref{eq:defDz}). Substituting them into
the expression (\ref{eq:defB+}), we obtain
\begin{align}
B^{+}(z)=\tB^{+}(z;\alpha+\cN,\cN)+\frac{4(\alpha+\cN-1)D_{1}^{+}
 (z;\alpha+\cN,\cN)}{f(z;\alpha+\cN)},
\end{align}
with
\begin{align}
\lefteqn{
\tB^{+}(z;\alpha+\cN,\cN)=-\frac{B_{4}^{+}(z;\alpha+\cN,\cN)}{
 2\alpha+\cN-1}a_{4}-[(\alpha+2\cN)a_{3}-a_{2}]z^{2}}
 \hspace{30pt}\notag\\
&-\bigl\{[3\alpha^{2}+(5\cN-2)\alpha+2(\cN-1)^{2}]a_{3}+(\cN+3)a_{2}
 -a_{1}\bigr\}z\notag\\[5pt]
&+4(\alpha+\cN-1)(\alpha+1)a_{3}+(\alpha+\cN-1)(3\alpha+1)a_{2}
 -(\alpha+\cN+3)a_{1},\\
\lefteqn{
D_{1}^{+}(z;\alpha+\cN,\cN)=\frac{(\alpha+\cN-1)D_{14}^{+}
 (z;\alpha+\cN,\cN)}{2\alpha+\cN-1}a_{4}-[(\alpha^{2}+\cN\alpha+2\cN-2)
 a_{3}}\hspace{30pt}\notag\\
&+(2\alpha+\cN-3)a_{2}-a_{1}]z-(\alpha+\cN)(\alpha+\cN-1)(\alpha+1)a_{3}
 \notag\\[5pt]
&-(\alpha+\cN-1)(2\alpha+\cN-1)a_{2}+(\alpha+\cN)a_{1},
\end{align}
where
\begin{align}
\lefteqn{
B_{4}^{+}(z;\alpha+\cN,\cN)=2\cN(2\alpha+\cN-1)z^{3}-[3\alpha^{3}
 -9\alpha^{2}-2(2\cN^{\,2}+\cN-2)\alpha}\hspace{30pt}\notag\\
&-2\cN^{\,2}(\cN-1)]z^{2}
 -(\alpha+\cN-1)[7\alpha^{3}+7(\cN-1)\alpha^{2}+(2\cN^{\,2}-9\cN+16)
 \alpha\notag\\
&-2(\cN-1)(\cN-4)]z
 -(\alpha+\cN-1)[2\alpha^{4}+(3\cN-7)\alpha^{3}+(\cN^{\,2}-8\cN-11)
 \alpha^{2}\notag\\
&-(\cN^{\,2}+31\cN-36)\alpha-4(\cN-1)(3\cN-4)],
\label{eq:B+++}
\\
\lefteqn{
D_{14}^{+}(z;\alpha+\cN,\cN)=[\alpha^{3}+(\cN+3)\alpha^{2}+(11\cN-16)
 \alpha+4(\cN-1)(\cN-2)]z}\hspace{30pt}\notag\\
&+(\alpha+\cN)[\alpha^{3}+(\cN+3)\alpha^{2}+(8\cN-9)\alpha+(\cN-1)
 (3\cN-4)].
\label{eq:D1+++}
\end{align}
Similarly, substituting (\ref{eq:defAz})--(\ref{eq:defDz}) into
expression (\ref{eq:defC+}), we obtain
\begin{align}
C^{+}(z)=\tC^{+}(z;\alpha+\cN,\cN)+\frac{4D_{2}^{+}(z;\alpha+\cN,\cN)
 }{f(z;\alpha+\cN)},
\end{align}
with
\begin{align}
\lefteqn{
\tC^{+}(z;\alpha+\cN,\cN)=\cN(\cN+1)a_{4}z^{2}
 }\hspace{30pt}\notag\\[5pt]
&-(\cN+1)\biggl[\frac{3\alpha^{3}+3(\cN-3)\alpha^{2}-(\cN^{\,2}+4\cN-4)
 \alpha-\cN^{\,2}(\cN-1)}{2\alpha+\cN-1}a_{4}\notag\\
&-(\alpha+\cN)a_{3}+a_{2}\biggr]z+c_{0}^{+},\\
\lefteqn{
D_{2}^{+}(z;\alpha+\cN,\cN)=\frac{(\alpha+\cN)(\alpha+\cN-1)
 D_{24}^{+}(z;\alpha+\cN,\cN)}{2\alpha+\cN-1}a_{4}
 }\hspace{30pt}\notag\\
&-[(\alpha+\cN)(\alpha+\cN-1)(\alpha+1)a_{3}+(\alpha+\cN-1)
 (2\alpha+\cN-1)a_{2}-(\alpha+\cN)a_{1}]z\notag\\[5pt]
&-(\alpha+\cN-1)\bigl\{\alpha(\alpha+\cN)^{2}a_{3}+[2\alpha^{2}
 +3(\cN-1)\alpha+\cN^{\,2}-\cN+2]a_{2}\notag\\[5pt]
&-(\alpha+\cN)a_{1}\bigr\},
\end{align}
where $c_{0}^{+}$ is an irrelevant constant which depends on $\alpha$
and $\cN$, and $D_{24}^{+}(z;\alpha+\cN,\cN)$ is given by
\begin{align}
D_{24}^{+}(z;\alpha+\cN,\cN)=&\,[\alpha^{3}+(\cN+3)\alpha^{2}
 +(8\cN-9)\alpha+(\cN-1)(3\cN-4)]z\notag\\
&+(\alpha+\cN)(\alpha+\cN-1)[\alpha^{2}+3\alpha+2(\cN-1)].
\label{eq:D2+++}
\end{align}
{}From expressions (\ref{eq:chH+2})--(\ref{eq:D2+++}), it is
evident that the operator $\check{H}^{+}$ consists of four linearly
independent operators, each of which preserves the linear space
$\tcV_{\cN}^{(\rmX_{2b})}[z;\alpha+\cN]$. Setting all but one of
the parameters $a_{i}$ ($i=1,\dots,4$) to be $0$ and replacing
$\alpha+\cN$ by $\alpha$, we can extract the four second-order
linear differential operators which preserve the second exceptional
polynomial subspace $\tcV_{\cN}^{(\rmX_{2b})}[z;\alpha]$. The
first operator $K_{1}$ associated with $a_{1}$ is given by
\begin{align}
K_{1}=z\del_{z}^{\,2}+(z-\alpha-3)\del_{z}+\frac{4}{f(z;\alpha)}
 \left[(\alpha-1)(z+\alpha)\del_{z}+\alpha(z+\alpha-1)\right].
\end{align}
The action of $K_{1}$ on $\bar{\chi}_{n}(z;\alpha)$ ($n=1,2,3,
\ldots$) reads as
\begin{align}
K_{1}\bar{\chi}_{n}(z;\alpha)=(n+1)\bar{\chi}_{n}(z;\alpha)
 -(n-1)(\alpha-n-1)\bar{\chi}_{n-1}(z;\alpha).
\end{align}
Hence, it preserves not only $\tcV_{\cN}^{(\rmX_{2b})}[z;\alpha]$
for a specific value of $\cN$ but also an infinite flag of the
spaces
\begin{align}
\tcV_{1}^{(\rmX_{2b})}[z;\alpha]\subset\tcV_{2}^{(\rmX_{2b})}
 [z;\alpha]\subset\dots\subset\tcV_{\cN}^{(\rmX_{2b})}[z;\alpha]
 \subset\cdots.
\end{align}
Therefore, the operator $K_{1}$ is not only quasi-solvable but also
solvable. The second operator $K_{2}$ associated with the parameter
$a_{2}$ is given by
\begin{align}
K_{2}=&\,[z^{2}+(\alpha-1)(\alpha-\cN-1)]\del_{z}^{\,2}+[z^{2}-(\cN+3)z
 +(\alpha-1)(3\alpha-3\cN+1)]\del_{z}\notag\\
&-(\cN+1)z-\frac{4(\alpha-1)}{f(z;\alpha)}\bigl\{[(2\alpha-\cN-3)z
 +(\alpha-1)(2\alpha-\cN-1)]\del_{z}\notag\\
&+(2\alpha-\cN-1)z+2\alpha^{2}-(\cN+3)\alpha+2(\cN+1)\bigr\}.
\end{align}
The action of $K_{2}$ on $\bar{\chi}_{n}(z;\alpha)$ ($n=1,2,3,\ldots$)
reads as
\begin{align}
\lefteqn{
(\alpha-n+2)(\alpha-n+1)(\alpha-n)(\alpha-n-1)K_{2}\bar{\chi}_{n}
 (z;\alpha)=(n-\cN)(\alpha-n+2)
}\hspace{20pt}\notag\\
&\times(\alpha-n+1)^{2}(\alpha-n)\bar{\chi}_{n+1}(z;\alpha)
 -(\alpha-n+2)(\alpha-n+1)s^{+}(n,\alpha,\cN)\bar{\chi}_{n}(z;\alpha)
 \notag\\
&+(n-1)(\alpha-1)(\alpha-\cN-1)(\alpha-n-1)^{2}[3\alpha^{2}
 -6(n-1)\alpha+3n^{2}-6n+4]\bar{\chi}_{n-1}(z;\alpha)\notag\\
&+(n-1)(n-2)(\alpha-1)(\alpha-\cN-1)(\alpha-n)^{2}(\alpha-n-1)^{2}
 \bar{\chi}_{n-2}(z;\alpha),
\end{align}
where $s^{+}(n,\alpha,\cN)$ is given by
\begin{align}
s^{+}(n,\alpha,\cN)=&\,2\alpha^{3}-[n^{2}-(\cN-2)n-\cN+3]\alpha^{2}
 \notag\\
&+[2n^{3}-(2\cN+1)n^{2}-(3\cN+2)n+\cN+3]\alpha\notag\\
&-n^{4}+(\cN+1)n^{3}+(2\cN+3)n^{2}+(\cN+1)n-2(\cN+1).
\end{align}
Hence, it certainly satisfies $K_{2}\tcV_{\cN}^{(\rmX_{2b})}[z;\alpha]
\subset\tcV_{\cN}^{(\rmX_{2b})}[z;\alpha]$ only for a specific value of
$\cN\in\bbN$. The third operator $K_{3}$ associated with $a_{3}$ is
given by
\begin{align}
K_{3}=&\,(z+2\alpha-\cN-1)z^{2}\del_{z}^{\,2}-\bigl\{(\alpha+\cN)z^{2}
 +[3\alpha^{2}-(\cN+2)\alpha-2(\cN-1)]z\notag\\[5pt]
&-4(\alpha-1)(\alpha-\cN+1)\bigr\}\del_{z}+(\cN+1)\alpha z\notag\\
&-\frac{4(\alpha-1)}{f(z;\alpha)}\bigl\{[(\alpha^{2}-\cN\alpha+2\cN-2)z
 +\alpha(\alpha-1)(\alpha-\cN+1)]\del_{z}\notag\\
&+\alpha[(\alpha-\cN+1)z+\alpha(\alpha-\cN)]\bigr\}.
\end{align}
The action of $K_{3}$ on $\bar{\chi}_{n}(z;\alpha)$ admits a relatively
simple form as in the case of $J_{3}$ as follows:
\begin{align}
K_{3}\bar{\chi}_{n}(z;\alpha)=&-(n-\cN)(\alpha-n+1)\bar{\chi}_{n+1}
 (z;\alpha)-[(3n+1)\alpha^{2}\notag\\
&-(2n^{2}+\cN n+3\cN-2)\alpha+(\cN+1)n(n-1)]\bar{\chi}_{n}(z;\alpha),
\end{align}
and thus it also satisfies $K_{3}\tcV_{\cN}^{(\rmX_{2b})}[z;\alpha]
\subset\tcV_{\cN}^{(\rmX_{2b})}[z;\alpha]$ only for a specific value
of $\cN\in\bbN$. The fourth operator $K_{4}$ associated with the
parameter $a_{4}$ is given by
\begin{align}
\lefteqn{
(2\alpha-\cN-1)K_{4}=[(2\alpha-\cN-1)z+3\alpha^{2}-(3\cN+2)\alpha
 +\cN(\cN+1)]z^{3}\del_{z}^{\,2}}\hspace{30pt}\notag\\[5pt]
&-B_{4}^{+}(z;\alpha,\cN)\del_{z}+\cN(\cN+1)(2\alpha-\cN-1)z^{2}
 \notag\\[5pt]
&-(\cN+1)[3\alpha^{3}-3(2\cN+3)\alpha^{2}+2(\cN^{\,2}+7\cN+2)\alpha
 -4\cN(\cN+1)]z\notag\\
&+\frac{4(\alpha-1)}{f(z;\alpha)}\left[(\alpha-1)D_{14}^{+}
 (z;\alpha,\cN)\del_{z}+\alpha D_{24}^{+}(z;\alpha,\cN)\right],
\end{align}
where the functions $B_{4}^{+}(z;\alpha,\cN)$ and $D_{i4}^{+}
(z;\alpha,\cN)$ ($i=1,2$) are introduced in (\ref{eq:B+++}),
(\ref{eq:D1+++}), and (\ref{eq:D2+++}), respectively. The action of
$K_{4}$ on $\bar{\chi}_{n}(z;\alpha)$ ($n=1,2,3,\ldots$) reads as
\begin{align}
\lefteqn{
(\alpha-n-2)(\alpha-n-1)(\alpha-n)(\alpha-n+1)K_{4}\bar{\chi}_{n}
 (z;\alpha)=(n-\cN)(n-\cN+1)}\hspace{20pt}\notag\\
&\times(\alpha-n)^{2}(\alpha-n+1)^{2}\bar{\chi}_{n+2}(z;\alpha)
 +(n-\cN)(\alpha-n+1)^{2}t_{1}^{+}(n,\alpha,\cN)\bar{\chi}_{n+1}
 (z;\alpha)\notag\\
&+(\alpha-1)(\alpha-n-2)(\alpha-n-1)t_{2}^{+}(n,\alpha,\cN)
 \bar{\chi}_{n}(z;\alpha)+(n-1)\alpha(\alpha-1)\notag\\
&\times(\alpha-\cN-1)(\alpha-\cN)(\alpha-n-2)
 (\alpha-n-1)^{2}(\alpha-n)\bar{\chi}_{n-1}(z;\alpha),
\end{align}
and thus it surely preserves the space $\tcV_{\cN}^{(\rmX_{2b})}
[z;\alpha]$ for a specific value of $\cN\in\bbN$. In the above,
$t_{1}^{+}(n,\alpha,\cN)$ and $t_{2}^{+}(n,\alpha,\cN)$ are,
respectively, given by
\begin{align}
\lefteqn{
(2\alpha-\cN-1)t_{1}^{+}(n,\alpha,\cN)=3\alpha^{5}-(3n+6\cN+20)\alpha^{4}
 -[3n^{2}-(9\cN+26)n}\hspace{30pt}\notag\\
&-(2\cN^{\,2}+27\cN+50)]\alpha^{3}+[3n^{3}-4n^{2}-(3\cN^{\,2}+35\cN+52)n
 -7\cN^{\,2}-49\cN-52]\alpha^{2}\notag\\
&-[(3\cN+2)n^{3}-(7\cN+10)n^{2}-2(4\cN^{\,2}+21\cN+18)n-2(\cN+1)(5\cN+12)
 ]\alpha\notag\\
&+(\cN+1)[\cN n^{3}-(\cN+4)n^{2}-2(3\cN+4)n-4(\cN+1)]
\end{align}
and
\begin{align}
\lefteqn{
(2\alpha-\cN-1)t_{2}^{+}(n,\alpha,\cN)=(7n+1)\alpha^{5}-2[7n^{2}
 +(7\cN+3)n+2\cN-11]\alpha^{4}}\hspace{30pt}\notag\\
&+[7n^{3}+4(7\cN+4)n^{2}+(9\cN^{\,2}+7\cN-42)n+3\cN^{\,2}-21\cN+1]
 \alpha^{3}\notag\\
&-[(14\cN+11)n^{3}+2(9\cN^{\,2}+8\cN-10)n^{2}+(2\cN^{\,3}+\cN^{\,2}
 -29\cN-7)n\notag\\
&-(\cN+1)(11\cN-8)]\alpha^{2}+(\cN+1)(n-1)[(9\cN+4)n^{2}+(4\cN^{\,2}
 +7\cN-10)n\notag\\
&+2\cN(\cN+1)]\alpha-2\cN(\cN+1)^{2}n^{2}(n-1).
\end{align}
It is evident from the resulting actions of $K_{i}$ ($i=1,\dots,4$)
on $\bar{\chi}_{n}(z;\alpha)$ ($n=1,2,3,\ldots$) that the gauged
Hamiltonian $\check{H}^{+}$, and thus $\bar{H}^{+}$ as well, are
not only quasi-solvable but also solvable if and only if the condition
(\ref{eq:scond}), which is the solvability condition for $\tH^{-}$,
is satisfied. That is, $\tH^{-}$, $\check{H}^{+}$, and $\bar{H}^{+}$
get solvable only simultaneously.

The fact that the four operators $K_{i}$ ($i=1,\dots,4$) leave
the polynomial subspace $\tcV_{\cN}^{(\rmX_{2b})}[z;\alpha]$
invariant in spite of the existence of the fractional coefficients
$1/f(z;\alpha)$ is partially explained by factorization properties
of the polynomials $\bar{\chi}_{n}(z;\alpha)$ for all
$n=1,2,3,\ldots$ with the common factor $f(z;\alpha)$ under
the actions of two first-order linear differential operators
$O_{1}^{+}$ and $O_{2}^{+}$, which are analogous to
(\ref{eq:fact1}),
\begin{align}
O_{1}^{+}\bar{\chi}_{n}(z;\alpha)=-(\alpha-n-1)
 (\alpha-n)(\alpha-n+1)z^{n-1}f(z;\alpha)
\end{align}
and
\begin{align}
O_{2}^{+}\bar{\chi}_{n}(z;\alpha)=&\,[(\alpha-n)(\alpha-n+1)z^{2}
 +2(\alpha-n-1)(\alpha-n+1)(\alpha-1)z\notag\\
&+(n-1)(\alpha-n-1)(\alpha-n)(\alpha-1)]z^{n-2}f(z;\alpha),
\end{align}
where $O_{1}^{+}$ and $O_{2}^{+}$ are given by
\begin{align}
O_{1}^{+}=z\del_{z}-\alpha,\qquad
 O_{2}^{+}=(\alpha-1)\del_{z}+z+2\alpha-2.
\end{align}
It is easy to check that all the fractional parts having the factor
$1/f(z;\alpha)$ in $K_{i}$ ($i=1,\dots,4$) are expressed as linear
combinations of $O_{1}^{+}$ and $O_{2}^{+}$ and thus they map all
the polynomials $\bar{\chi}_{n}(z;\alpha)$ to other polynomials in $z$.
The latter fact is inevitable for the operators $K_{i}$ ($i=1,\dots,4$)
to preserve the polynomial subspace $\tcV_{\cN}^{(\rmX_{2b})}
[z;\alpha]$.

Finally, the $\cN$-fold SUSY pair of Hamiltonians $H^{\pm}$ in the
physical space is obtained by gauge transformations
\begin{align}
H^{\pm}=\rme^{-\cW_{\cN}^{\pm}}\bar{\tH}^{\pm}\rme^{\cW_{\cN}^{\pm}}
 \Bigr|_{z=z(q)},
\end{align}
where the gauge potentials $\cW_{\cN}^{\pm}$ are given by
\begin{align}
\cW_{\cN}^{\pm}(q)=\frac{\cN-1}{4}\ln|2A(z)|\pm\int\rmd z\,
 \frac{Q(z)}{2A(z)}\biggr|_{z=z(q)},
\label{eq:cWN+-}
\end{align}
and the change of variable $z(q)$ is determined by
\begin{align}
z'(q)^{2}=2A(z(q)).
\label{eq:zofq}
\end{align}
If we introduce two functions $E(q)$ and $W(q)$ in the physical space
as
\begin{align}
E(q)=\frac{z''(q)}{z'(q)},\qquad W(q)=-\frac{Q(z(q))}{z'(q)},
\label{eq:defEW}
\end{align}
following the other types of $\cN$-fold SUSY~\cite{Ta03a,GT04,GT05},
the $\cN$-fold SUSY Hamiltonians $H^{\pm}$ are expressed as
\begin{align}
H^{\pm}=&-\frac{1}{2}\frac{\rmd^{2}}{\rmd q^{2}}+\frac{1}{2}W(q)^{2}
 -\frac{\cN-1}{4}\left(E'(q)-\frac{\cN-1}{2}E(q)^{2}-2W'(q)
 -2E(q)W(q)\right)\notag\\
&-C(z(q))\pm\frac{\cN}{2}W'(q)+\frac{1\pm1}{2}\left(z''(q)
 \tw_{\cN-1}^{[\cN]}(z(q))+z'(q)^{2}\tw_{\cN-1}^{[\cN]\prime}(z(q))
 \right),
\label{eq:H+-}
\end{align}
where $C(z)$ and $\tw_{\cN-1}^{[\cN]}(z)$ are, respectively, given by
(\ref{eq:defCz}) and (\ref{eq:wN-1a}). The pair of gauge potentials
$\cW_{\cN}^{\pm}(q)$ in (\ref{eq:cWN+-}) is expressed in terms of
$E(q)$ and $W(q)$ as
\begin{align}
\cW_{\cN}^{\pm}(q)=\frac{\cN-1}{2}\int\rmd q\,E(q)\mp\int\rmd q\,
 W(q).
\label{eq:cWN+-2}
\end{align}
Similarly, the components of $\cN$-fold supercharges $P_{\cN}^{\pm}$
in the $q$-space are also obtained by the same gauge transformations,
\begin{align}
P_{\cN}^{\pm}=\rme^{-\cW_{\cN}^{\pm}}\bar{\tP}_{\cN}^{\pm}\,
 \rme^{\cW_{\cN}^{\pm}}\Bigr|_{z=z(q)}.
\end{align}
Making the repeated use of the identity
\begin{align}
\left(\frac{\rmd}{\rmd z}\pm\frac{f'(z;\alpha)}{f(z;\alpha)}\right)
 z'(q)^{-k}=z'(q)^{-k-1}\left(\frac{\rmd}{\rmd q}\pm F_{\alpha}(q)
 -k E(q)\right),
\end{align}
where the function $F_{\alpha}(q)$ is defined by
\begin{align}
F_{\alpha}(q)=\frac{f'(z(q);\alpha)}{f(z(q);\alpha)}z'(q),
\end{align}
we immediately have
\begin{align}
P_{\cN}^{-}=&\,\frac{f(z(q);\alpha)}{f(z(q);\alpha+\cN)}
 \prod_{k=0}^{\cN-1}\frac{f(z(q);\alpha+k+1)}{f(z(q);\alpha+k)}
 \biggl(\frac{\rmd}{\rmd q}+W(q)\notag\\
&-F_{\alpha+k+1}(q)+\frac{\cN-1-2k}{2}E(q)\biggr)
\label{eq:PN-}
\end{align}
and
\begin{align}
P_{\cN}^{+}=&\,\Biggl[\prod_{k=0}^{\cN-1}\left(\frac{\rmd}{\rmd q}
 -W(q)+F_{\alpha+\cN-k}(q)+\frac{\cN-1-2k}{2}E(q)\right)\notag\\
&\times\frac{f(z(q);\alpha+\cN-k)}{f(z(q);\alpha+\cN-k-1)}\Biggr]
 \frac{f(z(q);\alpha)}{f(z(q);\alpha+\cN)}.
\label{eq:PN+}
\end{align}
It is easy to check that they are connected by the correct relation
$P_{\cN}^{+}=(-1)^{\cN}(P_{\cN}^{-})^{\rmT}$. The algorithm
automatically ensures~\cite{GT05} that the obtained pair of
Hamiltonians $H^{\pm}$ in (\ref{eq:H+-}) and the $\cN$-fold
supercharges $P_{\cN}^{\pm}$ in (\ref{eq:PN-}) and (\ref{eq:PN+})
satisfy the intertwining relations
\begin{align}
P_{\cN}^{-}H^{-}=H^{+}P_{\cN}^{-},\qquad
 P_{\cN}^{+}H^{+}=H^{-}P_{\cN}^{+},
\end{align}
and thus, in particular, the two Hamiltonians $H^{-}$ and $H^{+}$ are
almost isospectral.

By the construction it is evident that the $\cN$-fold SUSY Hamiltonians
$H^{\pm}$ respectively preserve the vector spaces $\cV_{\cN}^{\pm}$
defined by
\begin{subequations}
\label{eqs:V-V+}
\begin{align}
\cV_{\cN}^{-}&=\tcV_{\cN}^{-}\,\rme^{-\cW_{\cN}^{-}}\Bigr|_{z=z(q)}
 =\braket{\tvph_{1}(z(q);\alpha),\dots,\tvph_{\cN}(z(q),\alpha)}\,
 \rme^{-\cW_{\cN}^{-}(q)},\\
\cV_{\cN}^{+}&=\bar{\cV}_{\cN}^{+}\,\rme^{-\cW_{\cN}^{+}}
 \Bigr|_{z=z(q)}
=\frac{\braket{\bar{\chi}_{1}(z(q);\alpha+\cN),\dots,\bar{\chi}_{\cN}
 (z(q);\alpha+\cN)}}{f(z(q);\alpha)f(z(q);\alpha+\cN)}\,
 \rme^{-\cW_{\cN}^{+}(q)}.
\end{align}
\end{subequations}
Hence, if, for instance, both $H^{-}$ and $H^{+}$ are Hermitian in
a Hilbert space $L^{2}(S)$ with $S\subset\bbR$, and $\cV_{\cN}^{-}$
and/or $\cV_{\cN}^{+}$ are subspaces of $L^{2}(S)$, then $H^{-}$
and/or $H^{+}$ are not only quasi-solvable but also quasi-exactly
solvable on $S$. In the latter cases, the solvable sectors
$\cV_{\cN}^{-}$ and/or $\cV_{\cN}^{+}$ provide parts of the
eigenfunctions of $H^{-}$ and/or $H^{+}$ defined in $L^{2}(S)$.

\section{Resulting $\cN$-fold SUSY Pairs of Quasi-solvable Potentials}
\label{sec:QSpot}

In this section, we shall construct explicitly $\cN$-fold SUSY
pairs of potentials in the physical $q$-space associated with
the two exceptional polynomial subspaces $\tcV_{\cN}^{(\rmX_{2a})}$
and $\tcV_{\cN}^{(\rmX_{2b})}$. As in the case of type B $\cN$-fold
SUSY~\cite{GT04}, they have no covariance under the linear
fractional transformations $GL(2,\bbC)$. Hence, we can at most
consider projective equivalence classes as was done in
Ref.~\cite{GKM07} for the $X_{1}$ subspace, or equivalently,
the type B monomial space; the forms of potentials are not invariant
under all projective transformations in general. In this paper, we
shall rather content ourselves with exhibiting a couple of particular
examples since the complete presentation of all the cases would
involve many complicated formulas. The functional types of potentials
such as rational, exponential, hyperbolic, and so on are determined
by the function $A(z)$ through Eq.~(\ref{eq:zofq}). In
Table~\ref{tb:class}, we just show the relations between the form of
$A(z)$ and the types of the potentials $V^{\pm}(q)$.
\begin{table}
\begin{center}
\tabcolsep=10pt
\begin{tabular}{ll}
\hline
$A(z)$ & Types of $V^{\pm}(q)$\\
\hline\hline
$a_{1}\neq0$, $a_{2}=a_{3}=a_{4}=0$ & \multirow{2}{120pt}{Rational}\\
$a_{4}\neq0$, $a_{1}=a_{2}=a_{3}=0$ & \\
\hline
$a_{1}a_{2}\neq0$, $a_{3}=a_{4}=0$ & Exponential\\
\hline
$a_{2}\neq0$, $a_{1}=a_{3}=a_{4}=0$ & \\
$a_{3}\neq0$, $a_{1}=a_{2}=a_{4}=0$ & Trigonometric or Hyperbolic\\
$a_{3}a_{4}\neq0$, $a_{1}=a_{2}=0$ & \\
\hline
Other cases & Elliptic\\
\hline
\end{tabular}
\caption{The relations between the forms of $A(z)$ and the types of
$V^{\pm}(q)$.}
\label{tb:class}
\end{center}
\end{table}
In this paper, we shall just consider the two simplest cases
where only $a_{1}$ (Example 1) or $a_{2}$ (Example 2) among
the four parameters introduced in (\ref{eq:tH-}) is non-zero.
In such restricted analyses, we can still reduce the freedom of
parameters by considering the scale transformation as was done in
Ref.~\cite{GT05} for the classification of type C $\cN$-fold SUSY
potentials. That is, by (\ref{eq:defAz}) and (\ref{eq:zofq}) a
rescaling of $a_{i}$ ($i=1,\dots,4$) and $c_{0}$ by an overall
non-zero constant $\nu$ affects the change of variable $z(q)$ as
\begin{align}
z(q;\nu a_{i},\nu c_{0})=z(\rnu q;a_{i},c_{0}).
\label{eq:sc1}
\end{align}
{}From this and Eqs.~(\ref{eq:cWN+-}) and (\ref{eq:defEW}),
we easily obtain the scaling relations:
\begin{subequations}
\begin{align}
E(q;\nu a_{i},\nu c_{0})&=\rnu E(\rnu q;a_{i},c_{0}),& F_{\alpha}
 (q;\nu a_{i},\nu c_{0})&=\rnu F_{\alpha}(\rnu q;a_{i},c_{0}),\\
W(q;\nu a_{i},\nu c_{0})&=\rnu W(\rnu q;a_{i},c_{0}),&
 \cW_{\cN}^{\pm}(q;\nu a_{i},\nu c_{0})&=\cW_{\cN}^{\pm}(\rnu q;
 a_{i},c_{0}),
\end{align}
\end{subequations}
Then, by formula (\ref{eq:H+-}) the potential terms are scaled as
\begin{align}
V^{\pm}(q;\nu a_{i},\nu c_{0})=\nu V^{\pm}(\rnu q;a_{i},c_{0}).
\label{eq:sc3}
\end{align}
Hence, we can fix the value of parameters $a_{1}$ or $a_{2}$ without
any loss of generality. In what follows, we shall exhibit for each
case the change of variable $z=z(q)$, the functions $E(q)$, $W(q)$,
and $\cW_{\cN}^{\pm}(q)$ which determine the gauge factors,
the pair of potentials $V^{\pm}(q)$, and the pair of solvable
sectors $\cV_{\cN}^{\pm}$. They are obtained by the calculations
of (\ref{eq:zofq}), (\ref{eq:defEW}), (\ref{eq:cWN+-}) or
(\ref{eq:cWN+-2}), (\ref{eq:H+-}), and (\ref{eqs:V-V+}),
respectively.\\

\noindent
\textit{Example 1.} $A(z)=2z$ [$a_{1}=2$].\\

\noindent
Change of variable: $z(q)=q^{2}$.\\

\noindent
Gauge factors:
\begin{align}
E(q)=\frac{1}{q},\qquad W(q)=q-\frac{2\alpha+\cN-8}{2q}
 -\frac{4(\alpha-1)(q^{2}+\alpha)}{f(q^{2};\alpha)\,q},\\
\cW_{\cN}^{\pm}(q)=\mp\frac{q^{2}}{2}\pm\frac{2\alpha+\cN\pm\cN\mp 1
 }{2}\ln|q|\mp\ln|f(q^{2};\alpha)|.
\end{align}
Potentials:
\begin{subequations}
\label{eqs:pots1}
\begin{align}
V^{-}(q)=&\,\frac{q^{2}}{2}+\frac{4\alpha^{2}-1}{8q^{2}}
 +4\left[\frac{q^{2}-\alpha+1}{f(q^{2};\alpha)}
 -\frac{4(\alpha-1)q^{2}}{f(q^{2};\alpha)^{2}}\right]-\alpha+3
 -c_{0},
\label{eq:pot1-}\\
V^{+}(q)=&\,\frac{q^{2}}{2}+\frac{4(\alpha+\cN)^{2}-1}{8q^{2}}
 +4\left[\frac{q^{2}-\alpha-\cN+1}{f(q^{2};\alpha+\cN)}
 -\frac{4(\alpha+\cN-1)q^{2}}{f(q^{2};\alpha+\cN)^{2}}\right]\notag\\
&-\alpha+\cN+3-c_{0}.
\end{align}
\end{subequations}
Solvable sectors:
\begin{subequations}
\label{eqs:solv1}
\begin{align}
\cV_{\cN}^{-}&=\braket{\tvph_{1}(q^{2};\alpha),\dots,\tvph_{\cN}(q^{2};
 \alpha)}\frac{q^{\alpha+1/2}\rme^{-q^{2}/2}}{f(q^{2};\alpha)},\\
\cV_{\cN}^{+}&=\braket{\bar{\chi}_{1}(q^{2};\alpha+\cN),\dots,
 \bar{\chi}_{\cN}(q^{2};\alpha+\cN)}\frac{q^{-\alpha-\cN+1/2}
 \rme^{q^{2}/2}}{f(q^{2};\alpha+\cN)}.
\end{align}
\end{subequations}
Since the discriminant of $f(z;\alpha)$ is $1-\alpha$, both of the
potentials $V^{\pm}(q)$ only have a common unique pole at $q=0$ for
$\alpha>1$. Hence, the $\cN$-fold SUSY system is naturally defined
on the half line $q\in S=(0,\infty)$. On the latter domain $S$,
it is evident from (\ref{eqs:solv1}) that $\cV_{\cN}^{-}(S)\subset
L^{2}(S)$ and $\cV_{\cN}^{+}(S)\not\subset L^{2}(S)$. Therefore,
it manifests unbroken $\cN$-fold SUSY of the system.
In addition, the solvability condition (\ref{eq:scond}) is
satisfied in the system, and thus the Hamiltonian $H^{-}$
preserves the infinite flag of the subspaces of $L^{2}(S)$,
\begin{align}
\cV_{1}^{-}(S)\subset\cV_{2}^{-}(S)\subset\cdots\subset
 \cV_{\cN}^{-}(S)\subset\cdots\subset L^{2}(S).
\label{eq:infla-}
\end{align}
Hence, the Hamiltonian $H^{-}$ is not only quasi-solvable but also
\emph{exactly solvable} on $S$ provided that the infinite flag
constitutes a complete set of the Hilbert space $L^{2}(S)$, namely,
\begin{align}
\overline{\cV_{\cN}^{-}(S)}\to L^{2}(S)\qquad(\cN\to\infty).
\label{eq:compl}
\end{align}
The fulfilment of the solvability condition (\ref{eq:scond}) also
guarantees that the other Hamiltonian $H^{+}$ preserves the infinite
flag of the spaces $\cV_{\cN}^{+}$ ($\cN=1,2,3,\ldots$) which are
not subspaces of $L^{2}(S)$,
\begin{align}
\cV_{1}^{+}(S)\subset\cV_{2}^{+}(S)\subset\cdots\subset\cV_{\cN}^{+}
 (S)\subset\cdots\not\subset L^{2}(S).
\label{eq:infla+}
\end{align}
On the other hand, Eqs.~(\ref{eqs:pots1}) tell us that
the other Hamiltonian $H^{+}$ has the same form as its partner
Hamiltonian $H^{-}$.
Indeed, $H^{+}$ is identical, up to an additive constant, to
$H^{-}$ with its parameter $\alpha$ replaced by $\alpha+\cN$.
Hence, the $\cN$-fold SUSY system has shape invariance for all
$\cN\in\bbN$. Combining it with the fact that $H^{\pm}
\cV_{\cN}^{\pm}\subset\cV_{\cN}^{\pm}$ for all $\cN=1,2,3,\ldots$,
we come to the conclusion that both $H^{-}$ and $H^{+}$ preserve
the two different infinite flags (\ref{eq:infla-}) and
(\ref{eq:infla+}) with suitable parameters. In particular, $H^{+}$
is also exactly solvable on $S$ if the completeness (\ref{eq:compl})
is assured, although its $\cN$-fold SUSY sector $\cV_{\cN}^{+}$
does not belong to the Hilbert space $L^{2}(S)$.

Finally, we note that the potential $2V^{-}(q)$ coincides, up to
the scaling factor $\nu=\omega/2$ determining the scaling relations
(\ref{eq:sc1})--(\ref{eq:sc3}), with case II rational radial
oscillator potential $V^{(-)}(x)$ in Ref.~\cite{Qu09}, Eq.~(2.17),
of which the eigenfunctions are expressed in terms of the
$X_{2}$-Laguerre polynomials of the second kind
$\tL_{2,n+2}^{(\alpha)}(z)$.\\

\noindent
\textit{Example 2.} $A(z)=(z^{2}+\zeta^{2})/2$
 [$a_{2}=1/2$, $\zeta^{2}=(\alpha-1)(\alpha+\cN-1)>0$].\\

\noindent
Change of variable: $z(q)=\zeta\sinh q$.\\

\noindent
Gauge factors:
\begin{align}
E(q)=&\,\tanh q,\\
W(q)=&\,\frac{\zeta}{2}\cosh q+\frac{3}{2}\tanh q
 +\frac{(\alpha-1)(\alpha+\cN-3)}{\zeta\cosh q}\notag\\
&+\left[(2\alpha+\cN-3)\zeta\sinh q+(\alpha-1)(2\alpha+\cN-1)\right]
 \frac{2(\alpha-1)}{f(\zeta\sinh q;\alpha)\,\zeta\cosh q},\\
\cW_{\cN}^{\pm}(q)=&\mp\frac{\zeta}{2}\sinh q\mp\zeta\gd q
 +\frac{\cN-1\pm1}{2}\ln|\cosh q|\mp\ln|f(\zeta\sinh q;\alpha)|.
\end{align}
Potentials:
\begin{align}
V^{-}(q)=&\,\frac{\zeta^{2}}{8}\cosh^{2}q+\frac{\cN-1}{4}\zeta
 \sinh q+\frac{4\alpha^{2}+4(\cN-4)\alpha+\cN^{\,2}+16}{8}-c_{0}
 \notag\\
&+\frac{1}{8\cosh^{2}q}\left[
 4(\cN-1)\zeta\sinh q+4\alpha^{2}+4(\cN-2)\alpha-\cN^{\,2}-2\cN
 +4\right]\notag\\
&-2(\alpha-1)\left[\frac{\zeta\sinh q-\alpha-\cN+3}{f(\zeta\sinh q;
 \alpha)}-2(\alpha-1)\frac{2\zeta\sinh q-\cN+1}{f(\zeta\sinh q;
 \alpha)^{2}}\right],
\end{align}
\begin{align}
V^{+}(q)=&\,\frac{\zeta^{2}}{8}\cosh^{2}q+\frac{3\cN-1}{4}\zeta
 \sinh q+\frac{4\alpha^{2}+4(\cN-4)\alpha+\cN^{\,2}+16}{8}-c_{0}
 \notag\\
&-\frac{1}{8\cosh^{2}q}\left[
 4(\cN+1)\zeta\sinh q-4\alpha^{2}-4(\cN-2)\alpha+\cN^{\,2}+6\cN
 -4\right]\notag\\
&-2(\alpha+\cN-1)\left[\frac{\zeta\sinh q-\alpha+3}{f(\zeta\sinh q;
 \alpha+\cN)}-2(\alpha+\cN-1)\frac{2\zeta\sinh q+\cN+1}{
 f(\zeta\sinh q;\alpha+\cN)^{2}}\right].
\end{align}
Solvable sectors:
\begin{subequations}
\label{eqs:solv2}
\begin{align}
\cV_{\cN}^{-}&=\braket{\tvph_{1}(\zeta\sinh q;\alpha),\dots,
 \tvph_{\cN}(\zeta\sinh q;\alpha)}\frac{\rme^{-\zeta(\sinh q)/2
 -\zeta\gd q}}{(\cosh q)^{\cN/2-1}f(\zeta\sinh q;\alpha)},\\
\cV_{\cN}^{+}&=\braket{\bar{\chi}_{1}(\zeta\sinh q;\alpha+\cN),\dots,
 \bar{\chi}_{\cN}(\zeta\sinh q;\alpha+\cN)}\frac{\rme^{\zeta(\sinh q)/2
 +\zeta\gd q}}{(\cosh q)^{\cN/2}f(\zeta\sinh q;\alpha+\cN)}.
\end{align}
\end{subequations}
In the above, $\gd q=\arctan(\sinh q)$ is the Gudermann function.
The Hamiltonians are only quasi-solvable but not solvable since
$a_{2}\neq0$ and the solvability condition (\ref{eq:scond}) is
not satisfied. The resulting potentials are of the generalized
P\"{o}shl--Teller types.

For $\alpha>1$, both of the potentials $V^{\pm}(q)$ have no
singularities at any finite $|q|$ and thus would be naturally defined
on the full real axis $\bbR$. However, neither $\cV_{\cN}^{-}(\bbR)$
nor $\cV_{\cN}^{+}(\bbR)$ belongs to the Hilbert space $L^{2}(\bbR)$,
which means that both $H^{-}$ and $H^{+}$ are only quasi-solvable but
are not quasi-exactly solvable on $\bbR$. Hence, $\cN$-fold SUSY of
the system is dynamically broken in this example.

For $\zeta^{2}<0$, that is, for $-\cN+1<\alpha<1$, the change of
variable is given by $z(q)=|\zeta|\cosh q$. The form of the potentials
in the latter case is similar to the above system for $\zeta^{2}>0$
with $\zeta$, $\cosh q$, and $\sinh q$ replaced by $|\zeta|$,
$\sinh q$, and $\cosh q$, respectively. However, there exists a pole at
$q=0$ in both of the potentials $V_{\cN}^{\pm}(q)$ irrespective of
the value of $\alpha$. In addition, at $z=1-\alpha(>0)$ we have
$f'(1-\alpha;\alpha)=0$ and $f(1-\alpha;\alpha)=\alpha-1(<0)$, which
means that the function $f(z;\alpha)$ has a positive root for all
$\alpha<1$. Hence, a natural domain of the system may be a half line
$S=(q_{0},\infty)$ with $f(|\zeta|\cosh q_{0};\alpha)=0$.

\section{Discussion and Summary}
\label{sec:discus}

In this paper, we have constructed a family of quasi-solvable and
$\cN$-fold SUSY quantum systems where each Hamiltonian preserves
an exceptional polynomial subspace of codimension $2$. We started
with the $X_{2}$ space $\tcV_{\cN}^{(\rmX_{2a})}[z;\alpha]$
and constructed the four linearly independent second-order
differential operators $J_{i}$ ($i=1,\dots,4$) which preserve it.
We then constructed the $\cN$-fold SUSY quantum systems by applying
the algorithm developed in Ref.~\cite{GT05}. As a by-product, we
have automatically obtained the other $X_{2}$ space
$\bar{\cV}_{\cN}^{(\rmX_{2b})}[z;\alpha]$ and the four linearly
independent second-order differential operators $K_{i}$
($i=1,\dots,4$) which preserve the latter space. This shows one of
the advantageous and powerful aspects of the framework of $\cN$-fold
SUSY. We presented the two particular examples of the $\cN$-fold
SUSY systems. The one is the pair of rational-type potentials
which coincide with the rational shape invariant potentials in
Ref.~\cite{Qu09} and thus are not only quasi-solvable but also
solvable. In addition, it turned out that they admit two linearly
independent analytic local solutions. The other is the pair of
hyperbolic-type potentials both of which are only quasi-solvable.
Dynamical $\cN$-fold SUSY breaking would take place in the second
example but not in the first example.

The polynomial parts of eigenfunctions of the rational potential
$V^{-}(q)$ in (\ref{eq:pot1-}) are, on the one hand, given by
the infinite flag of the spaces $\tcV_{\cN}^{(\rmX_{2a})}$ spanned
by the polynomials $\tvph_{n}(z;\alpha)$ in (\ref{eq:X2pol}) and
are, on the other hand, expressed in terms of the $X_{2}$-Laguerre
polynomials of the second kind $\tL_{2,\nu+2}^{(\alpha)}(z)$ in
Ref.~\cite{Qu09}. Hence, the former polynomials $\tvph_{n}(z;\alpha)$
and the latter $\tL_{2,\nu+2}^{(\alpha)}(z)$ must be connected by
linear transformations. In fact, the first few $X_{2}$-Laguerre
polynomials are expressed by linear combinations of the polynomials
$\tvph_{n}(z;\alpha)$ as follows:
\begin{align}
(\alpha-1)\tL_{2,2}^{(\alpha)}(z)&=\tvph_{1}(z;\alpha),\\
\alpha\tL_{2,3}^{(\alpha)}(z)&=-\tvph_{2}(z;\alpha)+(\alpha+2)
 \tvph_{1}(z;\alpha),\\
2(\alpha+1)\tL_{2,4}^{(\alpha)}(z)&=\tvph_{3}(z;\alpha)-2(\alpha+3)
 \tvph_{2}(z;\alpha)+(\alpha+2)(\alpha+3)\tvph_{1}(z;\alpha).
\end{align}
It also indicates that the polynomial system $\{\tL_{2,\nu+2}^{
(\alpha)}(z)\}_{\nu=0}^{\infty}$ would be obtained by the
Gram--Schmidt orthogonalization of the base system
$\{\tvph_{n}(z;\alpha)\}_{n=1}^{\infty}$ if it constitutes an
\emph{orthogonal} polynomial system with respect to a certain
inner product.

Similarly, the first few $X_{2}$-Laguerre polynomials of the
first kind $\tL_{1,\nu+2}^{(\alpha)}(z)$ in Ref.~\cite{Qu09}
are expressed by linear combinations of the polynomials
$\bar{\chi}_{n}(z;\alpha)$ in (\ref{eq:X2pol2}) as follows:
\begin{align}
\alpha(\alpha+1)\tL_{1,2}^{(\alpha)}(z)=&\,\bar{\chi}_{1}
 (-z;-\alpha),\\
(\alpha+1)(\alpha+2)\tL_{1,3}^{(\alpha)}(z)=&\,\bar{\chi}_{2}
 (-z;-\alpha)+(\alpha+3)\bar{\chi}_{1}(-z;-\alpha),\\
2(\alpha+2)(\alpha+3)\tL_{1,4}^{(\alpha)}(z)=&\,\bar{\chi}_{3}
 (-z;-\alpha)+2(\alpha+4)\bar{\chi}_{2}(-z;-\alpha)\notag\\
&+(\alpha+3)(\alpha+4)\bar{\chi}_{1}(-z;-\alpha).
\end{align}
Hence, we arrive at the following conjecture:
\begin{description}
\item[Conjecture 1] The $X_{2}$-Laguerre polynomial system of
the first kind $\{\tL_{1,\nu+2}^{(\alpha)}(z)\}_{\nu=0}^{\infty}$
would be obtained by the Gram--Schmidt orthogonalization of
the base system $\{\bar{\chi}_{n}(-z;-\alpha)\}_{n=1}^{\infty}$
with respect to a certain inner product, while the $X_{2}$-Laguerre
polynomial system of the second kind $\{\tL_{2,\nu+2}^{(\alpha)}
(z)\}_{\nu=0}^{\infty}$ would be obtained by the same
orthogonalization scheme of the base system $\{\tvph_{n}(z;\alpha)
\}_{n=1}^{\infty}$ with respect to another certain inner product.
\end{description}
Another important remaining issue is to establish a systematic
algorithm to calculate the characteristic polynomials $\cP_{\cN}$
of the superHamiltonian $\bH$ in the solvable sectors
$\cV_{\cN}^{\pm}$ which appear in the anti-commutators of
the $\cN$-fold supercharges $\bQ_{\cN}^{\pm}$~\cite{AS03},
\begin{align}
\bigl\{\bQ_{\cN}^{-},\bQ_{\cN}^{+}\bigr\}=2^{\cN}\cP_{\cN}(\bH).
\end{align}
In the cases of type A and C $\cN$-fold SUSYs, it was shown 
\cite{Ta03a,GT05} that they are given by the generalized
Bender--Dunne polynomials of critical degrees \cite{BD96} and
thus are systematically calculated via recursion relations.
It would be interesting to examine whether a similar approach
also works for the present $\cN$-fold SUSY systems.

One of possible continuations of the present work is to construct
$\cN$-fold SUSY associated with the $X_{2}$-Jacobi polynomials.
In this respect, we note that, in contrast to the $X_{2}$-Laguerre
cases where two different kinds were found, the two sets of
the Jacobi-type polynomial systems associated with the two
different extended Scarf I potentials were found to be identical
with each other~\cite{Qu09}. The latter fact together with
the present result that the two different kinds of the
$X_{2}$-Laguerre polynomial systems would be connected by
$\cN$-fold SUSY (cf., Eqs.~(\ref{eqs:V-V+})) leads to another
conjecture:
\begin{description}
\item[Conjecture 2] The pair of solvable sectors of $\cN$-fold SUSY
Hamiltonians $H^{\pm}$ associated with the $X_{2}$-Jacobi
polynomials $\tP_{1,\nu+2}^{(\alpha,\beta)}(z)$ are both spanned
by a one common polynomial system (with possibly different values
of parameters) which would generate the system
$\{\tP_{1,\nu+2}^{(\alpha,\beta)}(z)\}_{\nu=0}^{\infty}$ through
its Gram--Schmidt orthogonalization.
\end{description}
We note that the classification of $X_{2}$ subspaces has remained
unsolved yet. We expect that the present work, together with
investigation on quasi-solvability and $\cN$-fold SUSY associated
with $X_{2}$-Jacobi polynomials, would provide us crucial clues
to it.

Another possible research direction is to explore quasi-solvability
and $\cN$-fold SUSY associated with $X_{m}$ subspaces for $m>2$.
It would tell us how the shape invariant potentials associated
with the $X_{m}$ ($m>2$) polynomials in Refs.~\cite{Qu09,OS09}
are realized in the more general framework of $\cN$-fold SUSY.

\begin{acknowledgments}

We would like to thank B.~Bagchi for drawing our attention to
the paper~\cite{Qu09} which motivated us to do this work.
This work was partially supported by the National Cheng Kung
University under Grant No.\ HUA:98-03-02-227.

\end{acknowledgments}



\bibliography{refsels}
\bibliographystyle{npb}



\end{document}